\documentclass[12pt]{article}
\usepackage{a4wide,epsfig}

\renewcommand{\thefootnote}{\fnsymbol{footnote}}
\newcommand{\ice}[1]{\relax}

\newcommand{\beq}{\begin{equation}}

\newcommand{\eeq}{\end{equation}}

\newcommand{\gsim}{\;\rlap{\lower 3.5 pt \hbox{$\mathchar \sim$}} \raise 1pt
 \hbox {$>$}\;}
\newcommand{\lsim}{\;\rlap{\lower 3.5 pt \hbox{$\mathchar \sim$}} \raise 1pt
 \hbox {$<$}\;}

\begin{document}    

\title{\vskip-3cm{\baselineskip14pt
\centerline{\normalsize\hfill DESY 00--173}
\centerline{\normalsize\hfill Freiburg-THEP 00/17}
\centerline{\normalsize\hfill TTP00--25}
\centerline{\normalsize\hfill hep-ph/0012002}
\centerline{\normalsize\hfill December 2000}
}
\vskip.7cm
Three-loop non-diagonal current correlators in QCD\\
and\\
NLO corrections to single-top-quark production
}

\author{
{K.G. Chetyrkin}$^{a,b,}$\thanks{Permanent address:
Institute for Nuclear Research, Russian Academy of Sciences,
60th October Anniversary Prospect 7a, Moscow 117312, Russia.}
\,\,
and
{M. Steinhauser}$^c$
  \\[3em]
  { (a) Fakult{\"a}t f{\"u}r Physik,}\\
  { Universit{\"a}t Freiburg, D-79104 Freiburg, Germany }
  \\[.5em]
  { (b) Institut f\"ur Theoretische Teilchenphysik,}\\
  { Universit\"at Karlsruhe, D-76128 Karlsruhe, Germany}
  \\[.5em]
  { (c) II. Institut f\"ur Theoretische Physik,}\\ 
  { Universit\"at Hamburg, D-22761 Hamburg, Germany}
}
\date{}
\maketitle

\begin{abstract} 
\noindent 
The non-diagonal correlators of vector and
scalar currents are considered at three-loop order in QCD. The full
mass dependence is computed in the case where one of the quarks is
massless and the other one carries the mass $M$. As applications we
consider the single-top-quark production via the process $q\bar{q}\to
t\bar{b}$ and the decay rate of a charged Higgs into hadrons.
In both cases the computed NLO corrections are shown to be numerically 
much less important than the leading ones.

%

\end{abstract}

\thispagestyle{empty}
\newpage
\setcounter{page}{1}

\renewcommand{\thefootnote}{\arabic{footnote}}
\setcounter{footnote}{0}


\section{Introduction}
 
In high energy experiments where the center-of-mass energy,
$\sqrt{s}$, is much larger than the mass of the quarks, $M$, 
the latter can often be neglected or an expansion in $M/\sqrt{s}$
is sufficient to describe the experimental data.
However, there are situations where the full dependence on $M$
and $\sqrt{s}$ is required.
One can, e.g., think on the high precision which meanwhile has been
reached at LEP (CERN), SLC (SLAC) or TEVATRON (Fermilab)
or on situations where the center-of-mass energy is
of the same order of magnitude as the quark masses.
In particular for threshold phenomena the masses
are important and the full dependence is desirable.

A variety of important observables can be described by the correlators
of two currents with different tensorial structure. If the coupling of
the currents to quarks is diagonal quantities like $e^+ e^-$
annihilation into hadrons and the decay of the $Z$ boson are covered
by the vector and axial-vector current correlators. Total decay rates
of neutral 
CP even or CP odd Higgs bosons can be obtained with the help of the
scalar and pseudo-scalar current densities, respectively.  For these
cases the full mass dependence at order $\alpha_s^2$ has been computed
in~\cite{CheKueSte96} for the non-singlet and in~\cite{CheHarSte98} for
the singlet correlators.

In this work  we deal with   three-loop correlators of
currents which couple to two different quarks flavours: a massless
one and one carrying the  mass $M$.

The non-diagonal vector and axial-vector correlators describe
properties connected to the $W$ boson. The absorptive part of these
correlators is directly related to the decay width of a (highly
virtual) $W$ boson into quark pairs and gluons.  Of particular
interest in this connection is the single-top-quark production via the
process $q\bar{q}\to t\bar{b}$ \cite{CortPetr91,StelzerWillen95}. This
is because the process directly probes the matrix element $|V_{tb}|$
--- one of the few parameters of the Standard model not yet
experimentally measured. The imaginary part of the transversal $W$
boson polarization function constitutes a gauge invariant and finite
contribution of ${\cal O}(\alpha_s^2)$ to the process.

As an application of the (pseudo-)scalar current correlator we
consider the decay of a charged Higgs boson which occurs in extensions
of the Standard Model. The corrections provided in this paper describe
the total hadronic decay rate of such a  boson (coupled  to a massive
and a massless quark).

\ice{
The paper is organized as follows. In Section~\ref{sec:pade} we
provide useful definitions and describe the method we use for the
computation. The connection between the effective and full theory is
established in Section~\ref{sec:eff}. In Sections~\ref{sec:res}
and~\ref{sec:reseff} the results are discussed in the full and
effective theory and the final discussion and the conclusions are given
in Section~\ref{sec:con}.
}



\section{Definitions}

In the vector case the polarization function is defined through
\begin{eqnarray}
  \left(-q^2g_{\mu\nu}+q_\mu q_\nu\right)\,\Pi^v(q^2)
  +q_\mu q_\nu\,\Pi^v_L(q^2)
  &=&
  i\int {\rm d}x\,e^{iqx}\langle 0|Tj^v_\mu(x) j^{v\dagger}_\nu(0)|0 \rangle,
  \label{eq:pivadef}
\end{eqnarray}
with
$j_\mu^v = \bar{\psi}_1\gamma_\mu\psi_2$.
Only the transversal part $\Pi^v(q^2)$ will be considered in the following.
The definition of the scalar polarization function reads
\begin{eqnarray}
  q^2\,\Pi^s(q^2)
  &=&
  i\int {\rm d}x\,e^{iqx} \langle 0|Tj^s(x)j^{s\dagger}(0)|0 \rangle
  \,,
  \label{eq:pispdef}
\end{eqnarray}
with $j^s = (m(\mu)/M) \bar{\psi}_1\psi_2$ where $m(\mu)$ is the
$\overline{\rm MS}$ quark mass and $M$ is its pole mass. 
 Throughout this paper we consider
anti-commuting $\gamma_5$.  This is justified because only non-singlet
diagrams contribute ($\psi_1$ is supposed to be different from
$\psi_2$). Thus, within perturbation theory, the axial-vector and
pseudo-scalar correlators coincide with the vector and scalar ones,
respectively.

We consider  only the case where one of the quarks is
massive. Thus we will identify $\psi_1$ with $q$, the massless quark,
and $\psi_2$ with $Q$ which is supposed to be a heavy quark of mass $M$.
Furthermore it is convenient to introduce the dimensionless variable 
\begin{eqnarray}
  z &=& \frac{q^2}{M^2}
  \,,
\end{eqnarray}
where $M$ refers to the pole mass. 
For the overall normalization of $\Pi^\delta(q^2)$ ($\delta=v,s$)
we adopt the QED-like conditions $\Pi^\delta(0)=0$.

The physical observables $R(s)$ are related to $\Pi(q^2)$ through
\begin{eqnarray}
  R^v(s)
  &=& 12\pi \,\,\,\mbox{Im}\left[ \Pi^v(q^2=s+i\epsilon) \right]
  \,,
  \\
  R^s(s)
  &=&  8\pi \,\,\,\mbox{Im}\left[ \Pi^s(q^2=s+i\epsilon) \right]
  \,,
\end{eqnarray}
where the use of the variables
\begin{eqnarray}
  x\,\,=\,\,\frac{M}{\sqrt{s}}\,, && v\,\,=\,\, \frac{1-x^2}{1+x^2}\,,
\end{eqnarray}
turns out to be useful to describe the high energy and threshold
region, respectively.

The expansion of $\Pi^\delta$ in terms of $\alpha_s$ reads
($\delta=v,s$)
\begin{eqnarray}
  \Pi^\delta &=& \Pi^{(0),\delta} 
  + \frac{\alpha_s^{(n_f)}(\mu)}{\pi} C_F \Pi^{(1),\delta}
  + \left(\frac{\alpha_s^{(n_f)}(\mu)}{\pi}\right)^2 \Pi^{(2),\delta}
  + {\cal O}(\alpha_s^3)
  \,.
\label{eq:pidef}
\end{eqnarray}
It is convenient to decompose the three-loop term according to the
colour structure
\begin{eqnarray}
  \Pi^{(2),\delta} &=& C_F^2 \Pi_{FF}^{(2),\delta}
                    +  C_AC_F \Pi_{FA}^{(2),\delta}
                    +  C_FTn_l \Pi_{FL}^{(2),\delta}
                    +  C_FT \Pi_{FH}^{(2),\delta}
  \label{eq:pidef2}
  \,,
\end{eqnarray}
where analogous formulae hold for $R(s)$.
In Eq.~(\ref{eq:pidef2}) $\Pi_{FF}^{(2),\delta}$ corresponds to the
abelian part already present in QED whereas the non-abelian structure is
contained in $\Pi_{FA}^{(2),\delta}$. The remaining two structures
correspond to the fermionic contributions where $n_l$ counts the
number of massless quarks and $n_f=n_l+1$ is the total number of active
quark flavours.


\section{Calculation} 

A complete analytical computation of $\Pi^\delta(q^2)$ 
at three-loop order or its imaginary part is currently not feasible.
The method we use for the computation of the diagrams is based on
conformal mapping and Pad\'e 
approximation~\cite{FleTar94,pade,CheKueSte96,CheHarSte98}.
It allows for the computation of a semi-numerical approximation
for $\Pi^\delta(q^2)$.
The aim is the reconstruction of the function $\Pi^\delta(q^2)$ from the
knowledge of some moments for $z\to0$ and $z\to-\infty$ and 
additional partial information about the behaviour of
$R(s)$ close to the threshold, i.e. for $s\to M^2$.
Thus let us in the following briefly discuss the different kinematical
regions.

\subsection*{Expansions for small and large external momentum}

The number of diagrams which contribute at one-, two- and three-loop
order is relatively small and amounts to one, three and 31,
respectively.
Nevertheless we used {\tt
GEFICOM}~\cite{geficom} for the automatic computation.  {\tt GEFICOM}
uses {\tt QGRAF}~\cite{qgraf} for the generation of the diagrams. In
case an asymptotic expansion has to be applied {\tt
LMP}~\cite{Har:diss} or {\tt EXP}~\cite{Sei:dipl} are used for the
generation of the sub-diagrams. The occurring vacuum diagrams are
passed to {\tt MATAD}~\cite{matad} and the massless propagator type
diagrams are evaluated with {\tt MINCER}~\cite{mincer}. More details on
the automatic computation of Feynman diagrams can be found
in~\cite{HarSte98}.

We have been able to compute seven terms for small and eight terms
for large external momentum $q$ both for the vector and scalar correlators
of Eqs.~(\ref{eq:pivadef}) and~(\ref{eq:pispdef}), respectively.
This means expansion coefficients up to order $z^6$, respectively, $1/z^7$
are available for the Pad\'e procedure.
The analytical results are rather bulky and will be published elsewhere.

\subsection*{Threshold behaviour}

The Born results for the vector and scalar spectral functions
are proportional
to $v^2$ in the limit $v\to0$ as can be seen for the exact results
\begin{eqnarray}
  R^{(0),v}(s) &=& \frac{N_c}{2}
                   \left(1-x^2\right)^2
                   \left(2+x^2\right)
  \label{RvBorn}
  \,,
  \\
  R^{(0),s}(s) &=& N_c
                   \left(1-x^2\right)^2
  \label{RsBorn}
  \,.
\end{eqnarray}
Also the corrections
of order $\alpha_s$ to $R^v$ and $R^s$ are 
known~\cite{Schilcher:1981kr,Bro81,Chang:1982qq,ReiRubYaz81,Djouadi:1994ss}
and a similar threshold behaviour is observed: the 
spectral functions vanish like $v^2$ (modulo powers of $\ln v $).
This is valid in every order in $\alpha_s$
as follows from  Heavy Quark Effective Theory (HQET)~\cite{HQET}.
Actually the latter can be used 
to obtain the leading threshold behaviour of $R^v(s)$
and $R^s(s)$ at ${\cal O}(\alpha_s^2)$
from the corresponding correlators in HQET. In
particular, the renormalization group equation in the effective theory
is used to get the leading logarithmic behaviour at order
$\alpha_s^2$. Afterwards the decoupling
relation between the currents in the full
and effective theory~\cite{JiMus91,BroGro95,Gro98} is exploited
to get the information about $R^v(s)$ and $R^s(s)$.
For $\mu^2=M^2$ our results read
\begin{eqnarray}
R^{v,thr} &=& N_c v^2 \Bigg\{ 6 +
  \frac{\alpha_s^{(n_f)}(M)}{\pi}
  C_F\left(
    \frac{27}{2} 
    + 12\zeta_2 
    - 9\ln2
    - 9\ln v
  \right)
  \nonumber\\&&\mbox{}
  +\left(\frac{\alpha_s^{(n_f)}(M)}{\pi}\right)^2
  \Bigg[
  C_F^2\left(
    c^v_{FF}
   +
   \left(
    -\frac{147}{8} 
    - 30\zeta_2 
    + \frac{27}{2}\ln2
   \right)\ln v 
   + \frac{27}{4}\ln^2 v
  \right)
  \nonumber\\&&\mbox{}
  +C_AC_F\left(
    c^v_{FA}
   + 
   \left(
    -\frac{423}{8} 
    - 19\zeta_2 
    + \frac{33}{2}\ln2
   \right)\ln v 
   + \frac{33}{4} \ln^2 v
  \right)
  \nonumber\\&&\mbox{}
  +C_FTn_l\left(
    c^v_{FL}
   + 
   \left(
    \frac{39}{2} 
    + 8\zeta_2 
    - 6\ln2
    \right)\ln v 
    - 3 \ln^2 v
  \right)
  \nonumber\\&&\mbox{}
  +C_FT\left( 
   \frac{133}{8}
   -10\zeta_2
  \right)
  \Bigg]
  \Bigg\}
  \,,
  \label{eq:Rvthr}
  \\
R^{s,thr} &=& N_c v^2 \Bigg\{ 4 +
  \frac{\alpha_s^{(n_f)}(M)}{\pi}
  C_F\left(
    13 + 8\zeta_2 - 6\ln2 - 6\ln v
  \right)
  \nonumber\\&&\mbox{}
  +\left(\frac{\alpha_s^{(n_f)}(M)}{\pi}\right)^2
  \Bigg[
  C_F^2\left(
    c^s_{FF}
    + 
    \left(
    -\frac{73}{4} 
    - 20\zeta_2
    + 9\ln2
    \right) \ln v 
    + \frac{9}{2}\ln^2 v
  \right)
  \nonumber\\&&\mbox{}
  +C_AC_F\left(
    c^s_{FA}
    + 
   \left(
   -\frac{141}{4} 
   - \frac{38}{3}\zeta_2 
   + 11\ln2
   \right)\ln v 
   + \frac{11}{2}\ln^2 v
  \right)
  \nonumber\\&&\mbox{}
  +C_FTn_l\left(
  c^s_{FL}+
    \left(
     13 
     + \frac{16}{3}\zeta_2 
     - 4\ln 2
    \right) \ln v
    - 2\ln^2 v
  \right)
  \nonumber\\&&\mbox{}
  +C_FT\left(
    \frac{727}{36}
    - 12 \zeta_2
  \right)
  \Bigg]
  \Bigg\}
  \,,
  \label{eq:Rsthr}
\end{eqnarray}
with $\zeta_2=\pi^2/6$.
The constants $c^\delta_{FF}, c^\delta_{FA}$ and $c^\delta_{FL}$
($\delta=s,v$) are unknown. Note, however, that the leading term for
the colour structure $C_FT$ is completely determined.
In order to incorporate the available threshold information into out
method one
has to perform an analytical continuation of the expressions in 
Eqs.~(\ref{eq:Rvthr}) and~(\ref{eq:Rsthr}).
Taking the logarithmic parts of Eqs.~(\ref{eq:Rvthr}) and~(\ref{eq:Rsthr})
one obtains the quadratic and cubic logarithms for the polarization
functions.
They can be incorporated into the Pad\'e procedure~\cite{CheKueSte96}.


\section{\label{sec:res}Results and Applications}

\begin{figure}[t]
  \begin{center}
    \begin{tabular}{cc}
      \leavevmode
      \epsfxsize=7.cm
      \epsffile[110 280 470 560]{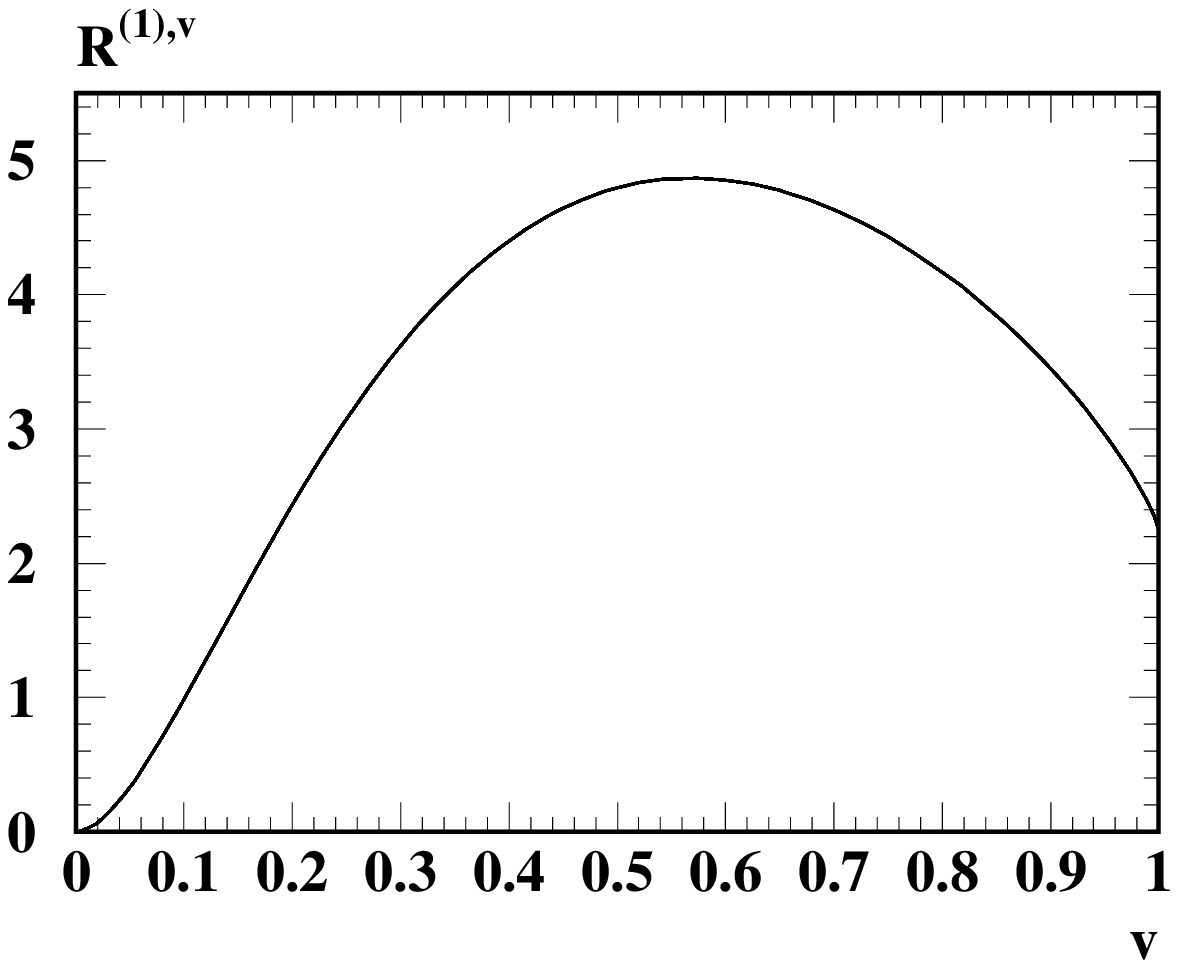}
      &
      \epsfxsize=7.cm
      \epsffile[110 280 470 560]{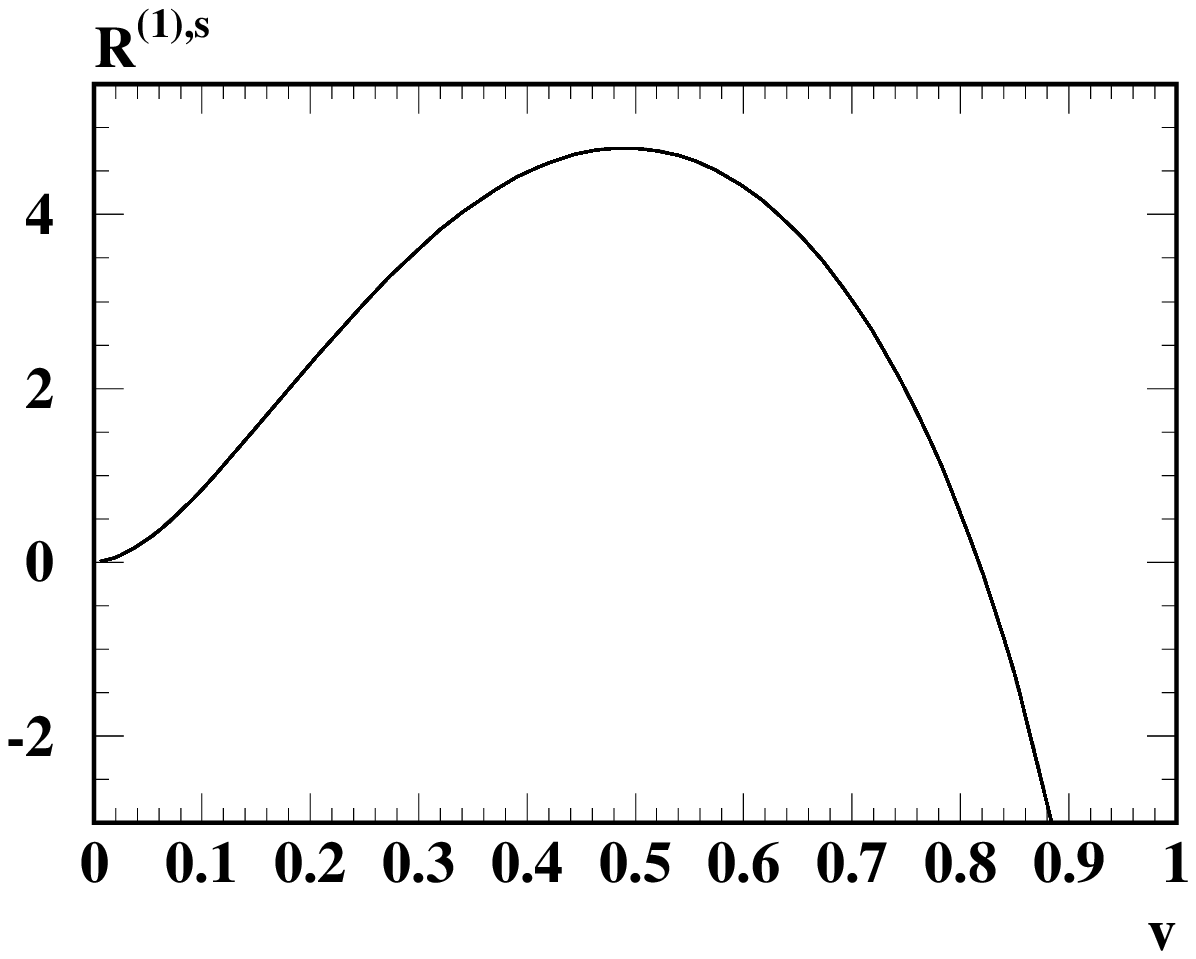}
    \end{tabular}
  \end{center}
  \caption{\label{fig:R2lv}$R^{(1),v}(s)$ and $R^{(1),s}(s)$ as a
    function of $v$.
          }
\end{figure}

In a first step we want to discuss the results for the 
spectral functions $R^v$ and $R^s$ which essentially corresponds to
the direct outcome of the Pad\'e method.
For the results we present
in the following, those Pad\'e approximants are chosen which contain for
their construction at least terms of order $z^5$ and $1/z^5$ 
in the small and large momentum region, respectively.
Furthermore we demand that the difference of the degree in the
numerator and denominator is less or equal to two.
In the numerical evaluations we specify $\mu^2=M^2$.

The analytical formulae which result from the semi-numerical Pad\'e
procedure are quite long. Thus we refrain from listing them
explicitly. Instead, a typical representative 
for each colour structure can be found under the URL
\verb|http://www-ttp.physik.uni-karlsruhe.de/Progdata/ttp00-25|.

\begin{figure}[t]
  \begin{center}
    \begin{tabular}{cc}
      \leavevmode
      \epsfxsize=7.cm
      \epsffile[110 280 470 560]{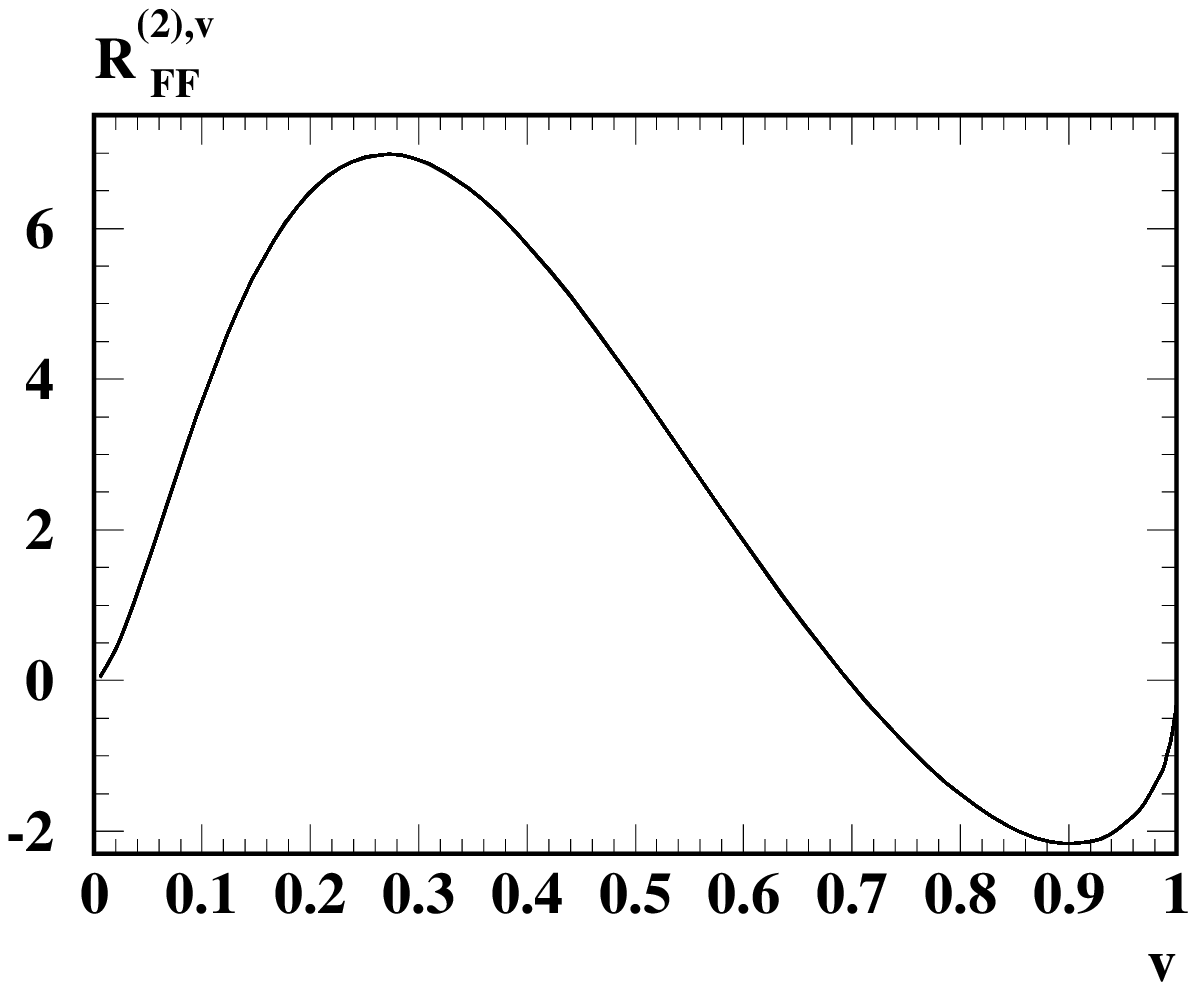}
      &
      \epsfxsize=7.cm
      \epsffile[110 280 470 560]{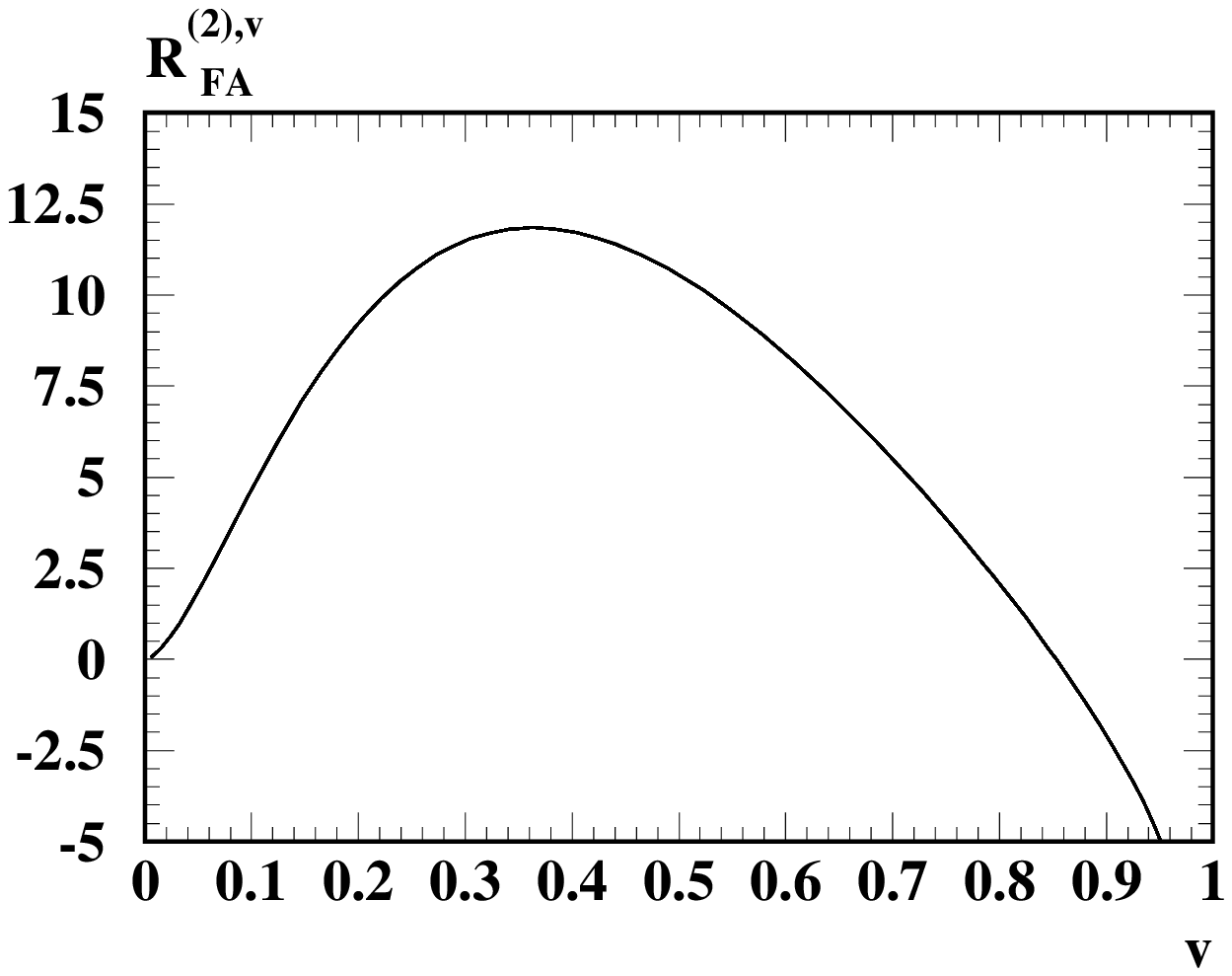}
      \\
      \epsfxsize=7.cm
      \epsffile[110 280 470 560]{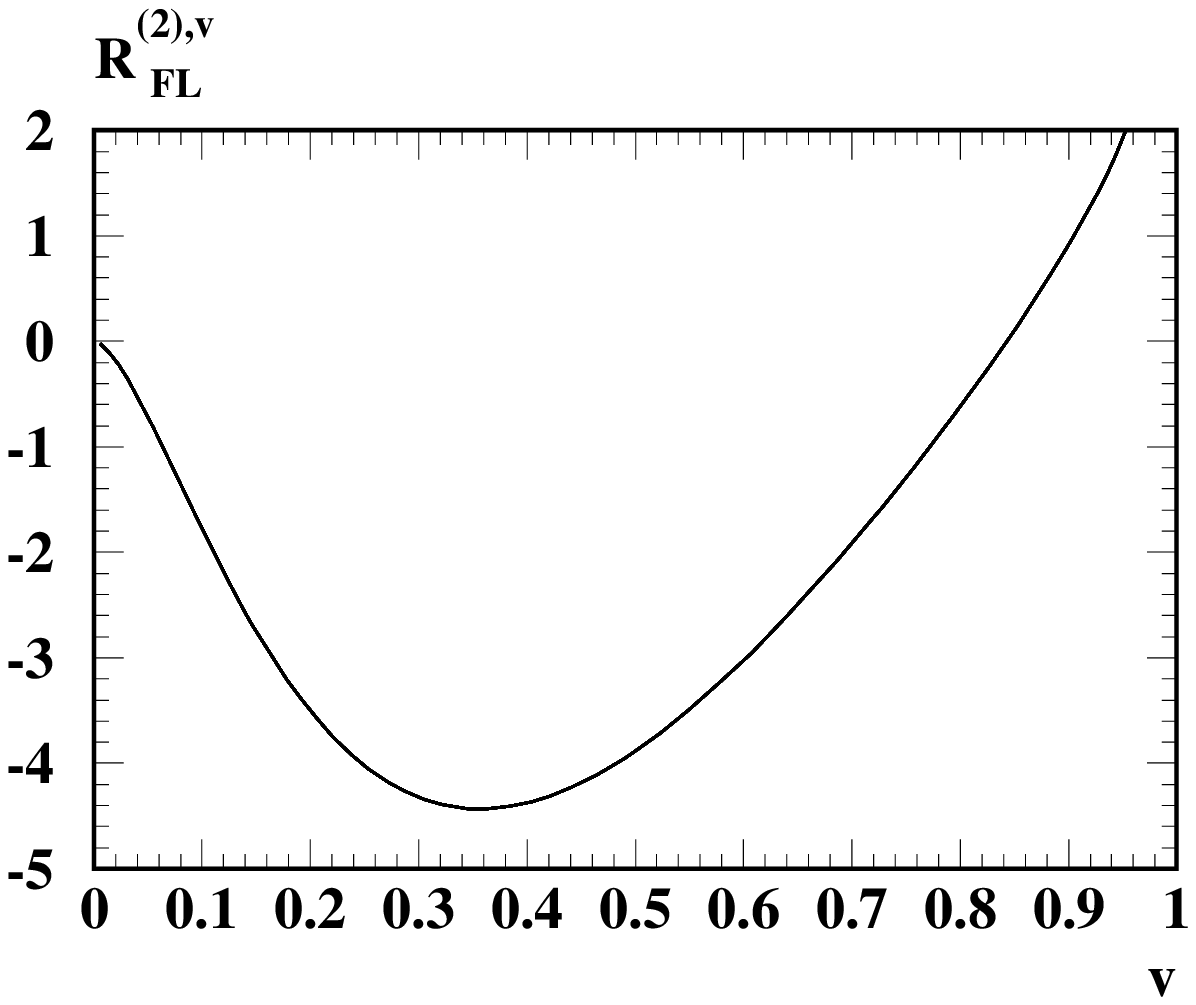}
      &
      \epsfxsize=7.cm
      \epsffile[110 280 470 560]{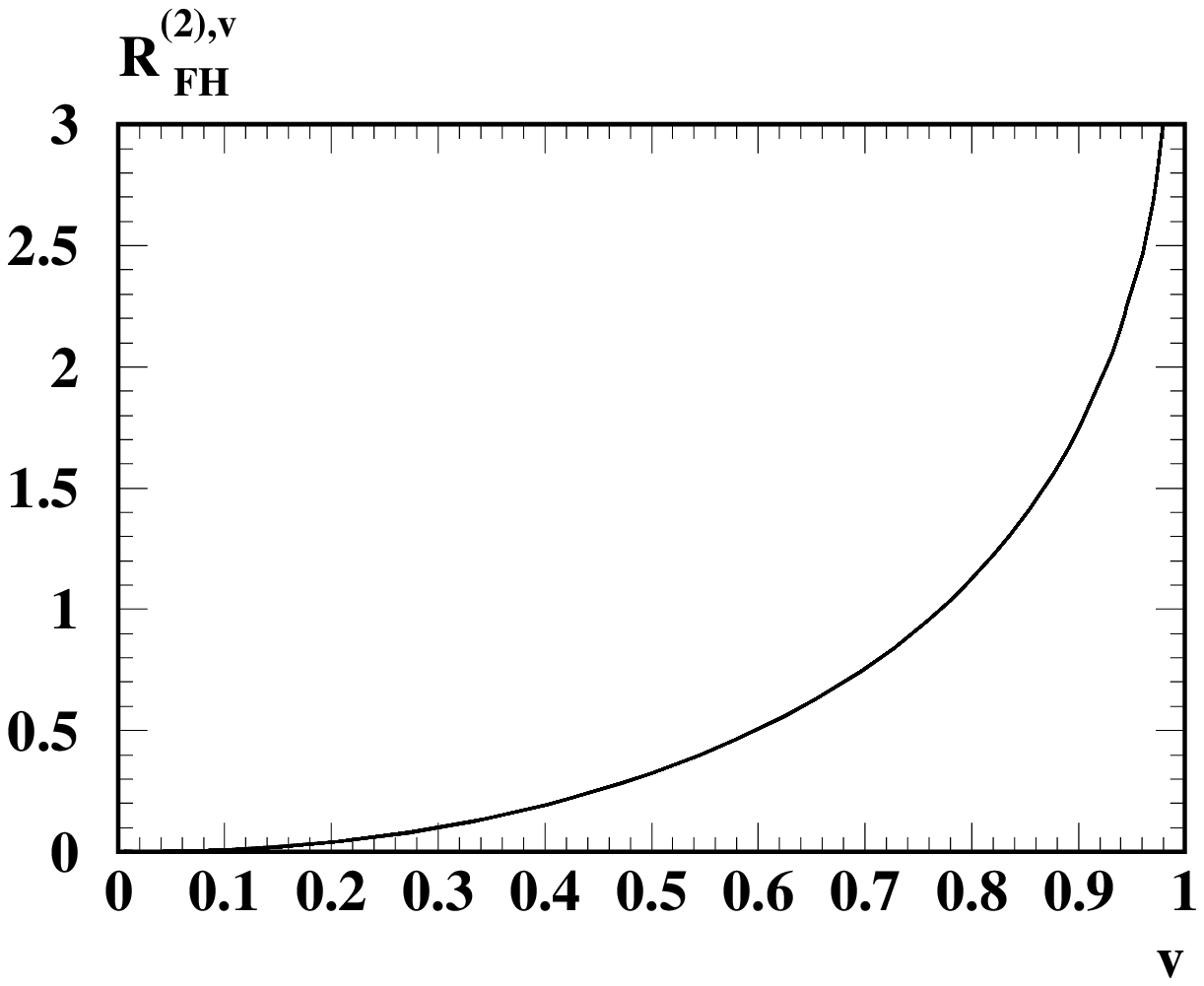}
    \end{tabular}
  \end{center}
  \caption{\label{fig:Rvv}$R^{(2),v}_{FF}(s)$, $R^{(2),v}_{FA}(s)$,
    $R^{(2),v}_{FL}(s)$  and $R^{(2),v}_{FH}(s)$ as a
    function of $v$.
          }
\end{figure}

\begin{figure}[t]
  \begin{center}
    \begin{tabular}{cc}
      \leavevmode
      \epsfxsize=7.cm
      \epsffile[110 280 470 560]{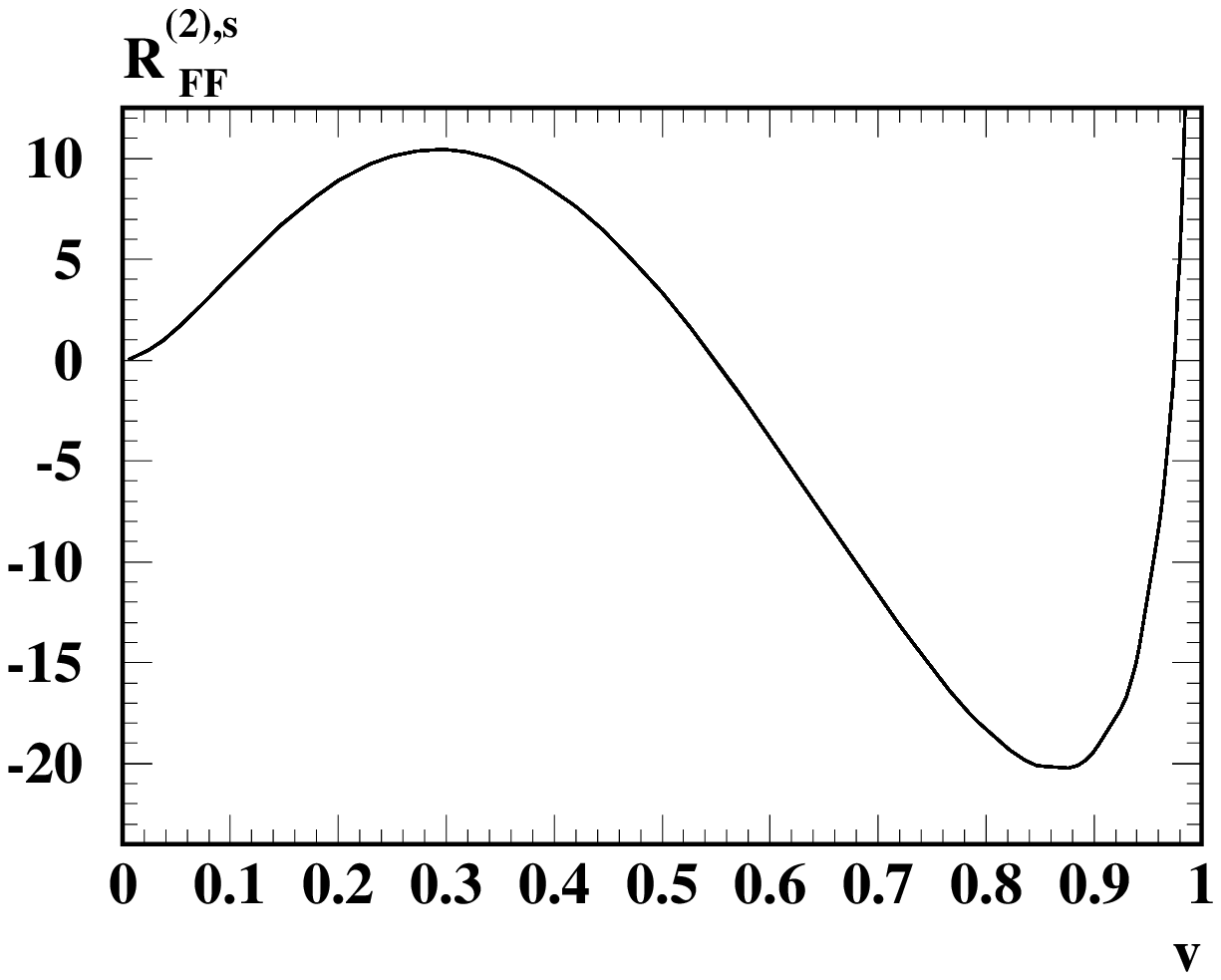}
      &
      \epsfxsize=7.cm
      \epsffile[110 280 470 560]{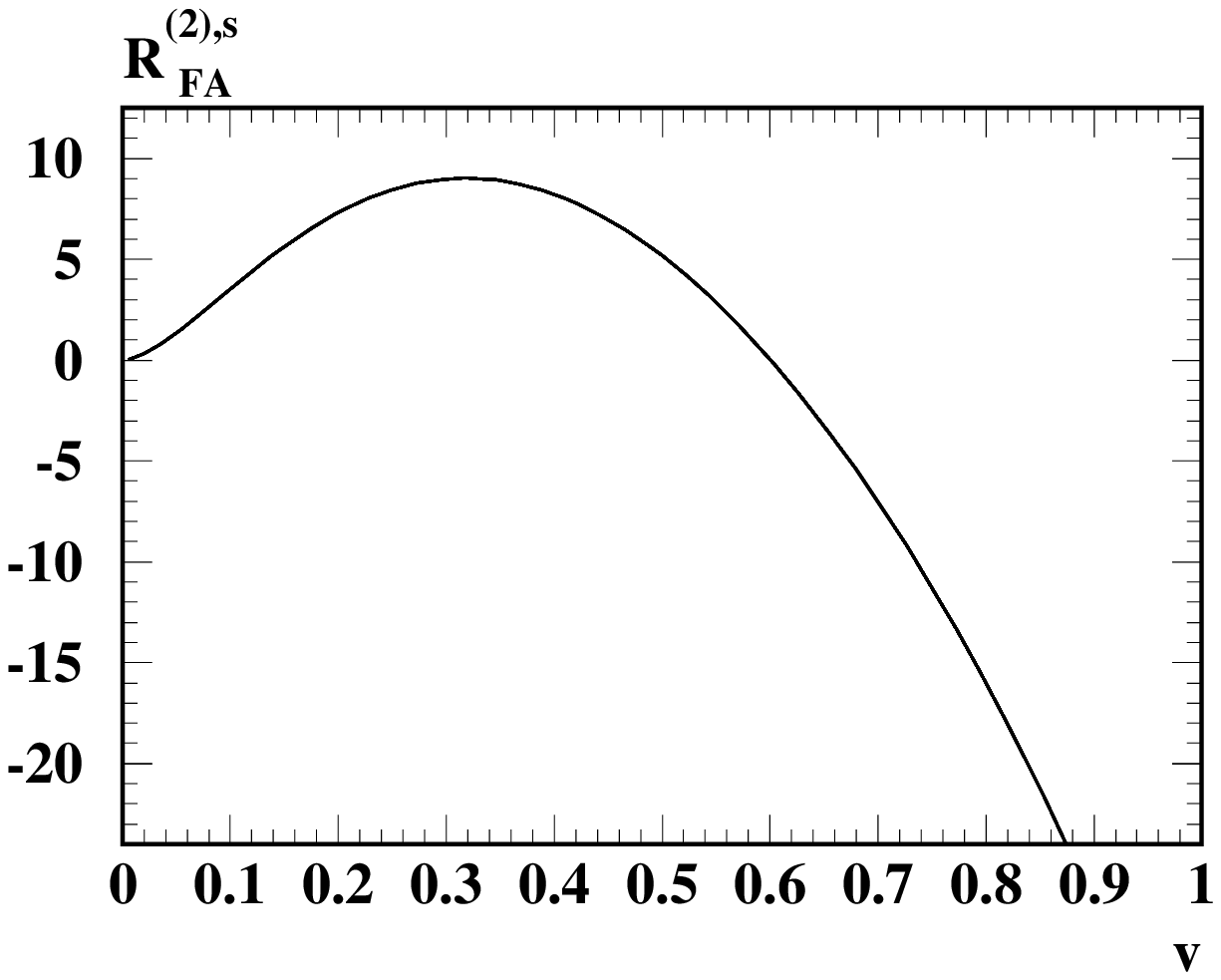}
      \\
      \epsfxsize=7.cm
      \epsffile[110 280 470 560]{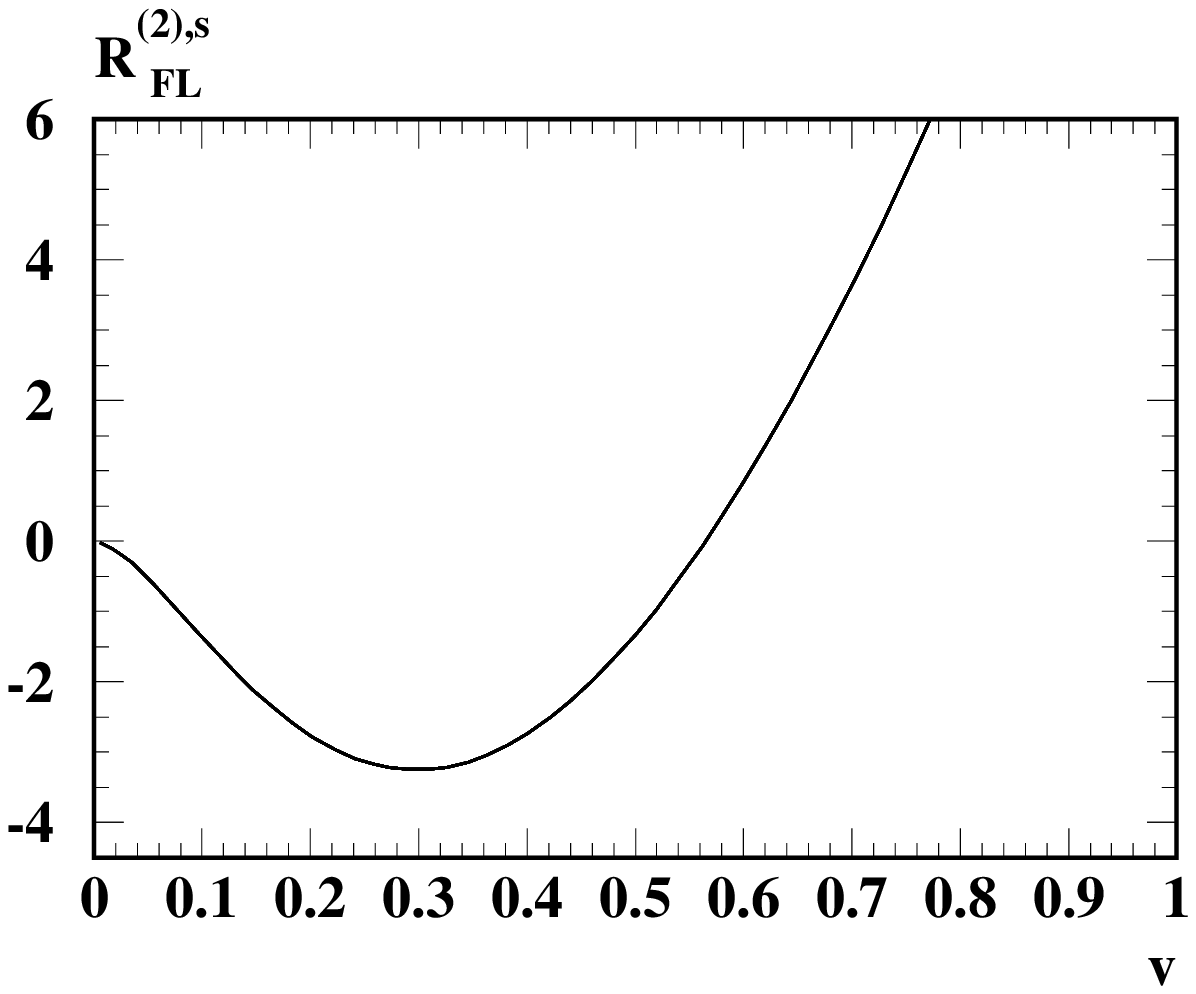}
      &
      \epsfxsize=7.cm
      \epsffile[110 280 470 560]{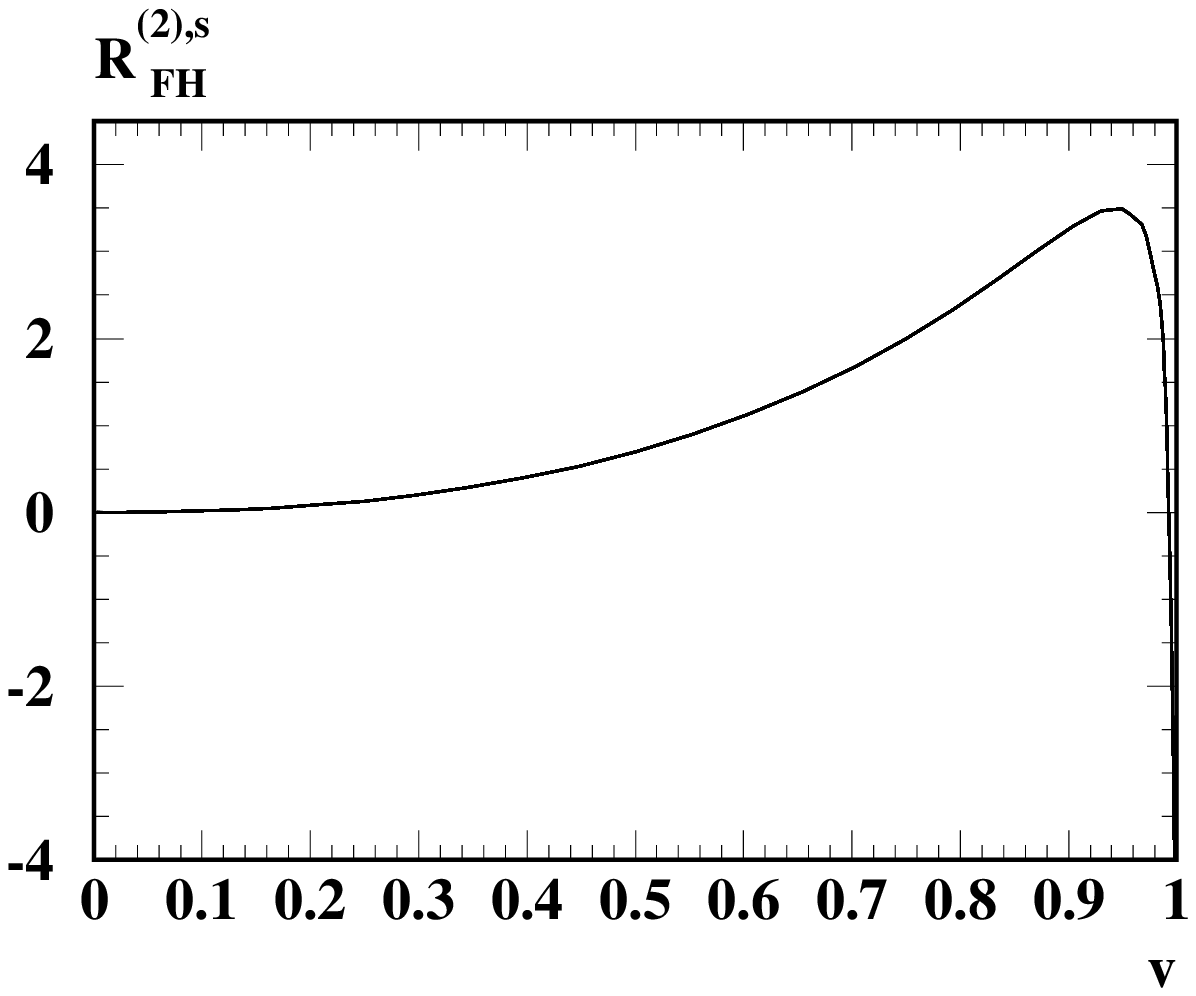}
    \end{tabular}
  \end{center}
  \caption{\label{fig:Rsv}$R^{(2),s}_{FF}(s)$, $R^{(2),s}_{FA}(s)$,
    $R^{(2),s}_{FL}(s)$  and $R^{(2),s}_{FH}(s)$ as a
    function of $v$.
          }
\end{figure}

At order $\alpha_s$ the exact results are known. Thus it is
instructive to make a comparison with the outcome of our
semi-numerical method. In Fig.~\ref{fig:R2lv} roughly 15
Pad\'e approximants
are plotted together with the exact curves. Note that even for small
values of $v$, where only the logarithmic parts of Eqs.~(\ref{eq:Rvthr})
and~(\ref{eq:Rsthr}) are incorporated in our method,
no difference both between the individual Pad\'e expression and the exact 
result is visible.
This is quite promising for the ${\cal O}(\alpha_s^2)$ results.
The corresponding curves are shown in 
Fig.~\ref{fig:Rvv} and Fig.~\ref{fig:Rsv} for the vector and scalar 
case, respectively. Again of the order of 15 Pad\'e approximants are
plotted and no difference is visible.
We should note that due to our choice $\mu^2=M^2$
in the high energy limit all curves tend to $\pm\infty$
except $R^{(2),v}_{FF}$ which reaches a constant for $v\to1$.
In particular, $R^{(2),s}_{FA}$ and $R^{(2),s}_{FL}$
tend to $+\infty$, respectively, $-\infty$.
This is not visible in the figures as the corresponding range in $v$
where the turn over takes place is
beyond the resolution in the plots.

\begin{figure}[t]
  \begin{center}
    \begin{tabular}{cc}
      \leavevmode
      \epsfxsize=7.cm
      \epsffile[110 280 470 560]{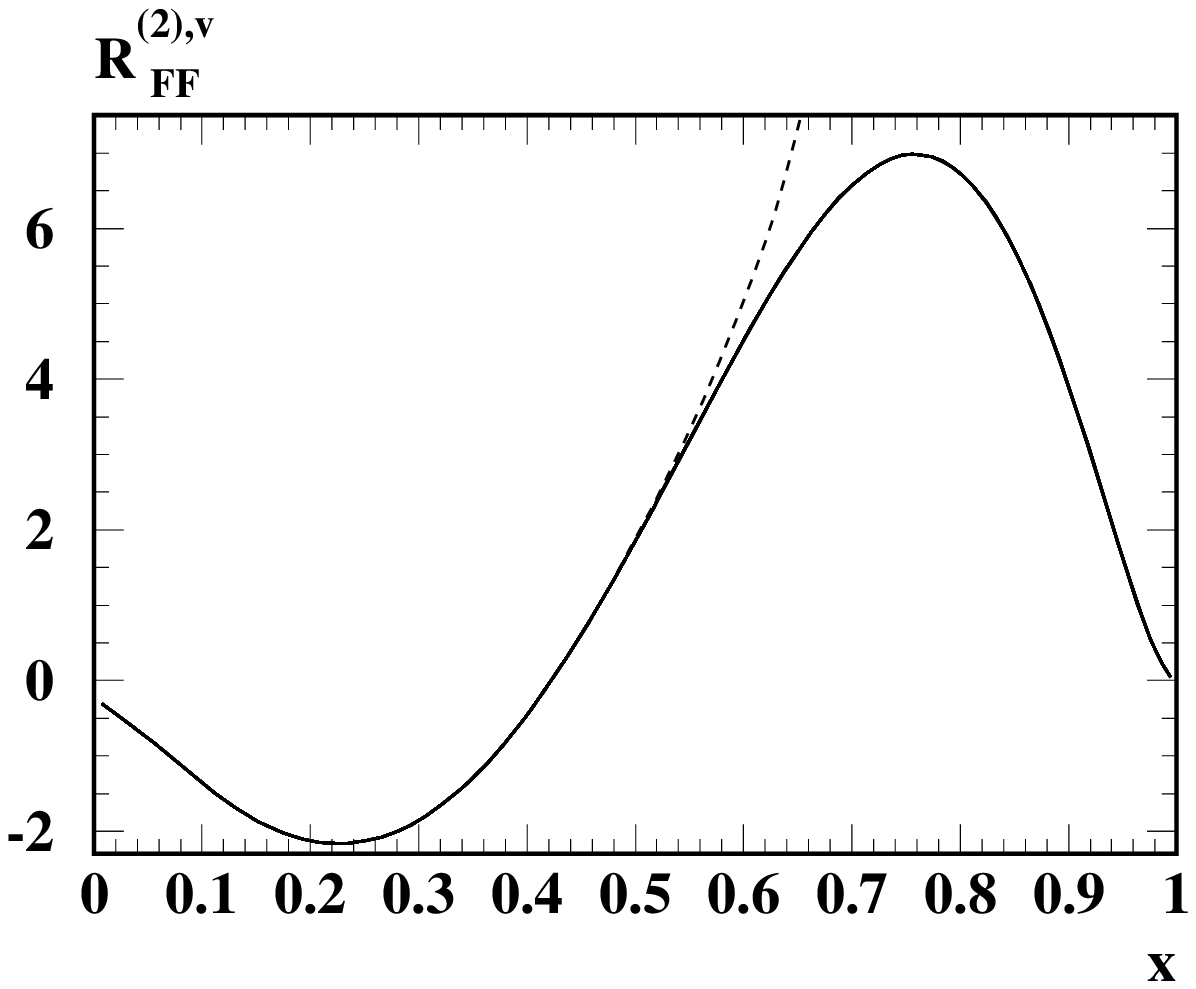}
      &
      \leavevmode
      \epsfxsize=7.cm
      \epsffile[110 280 470 560]{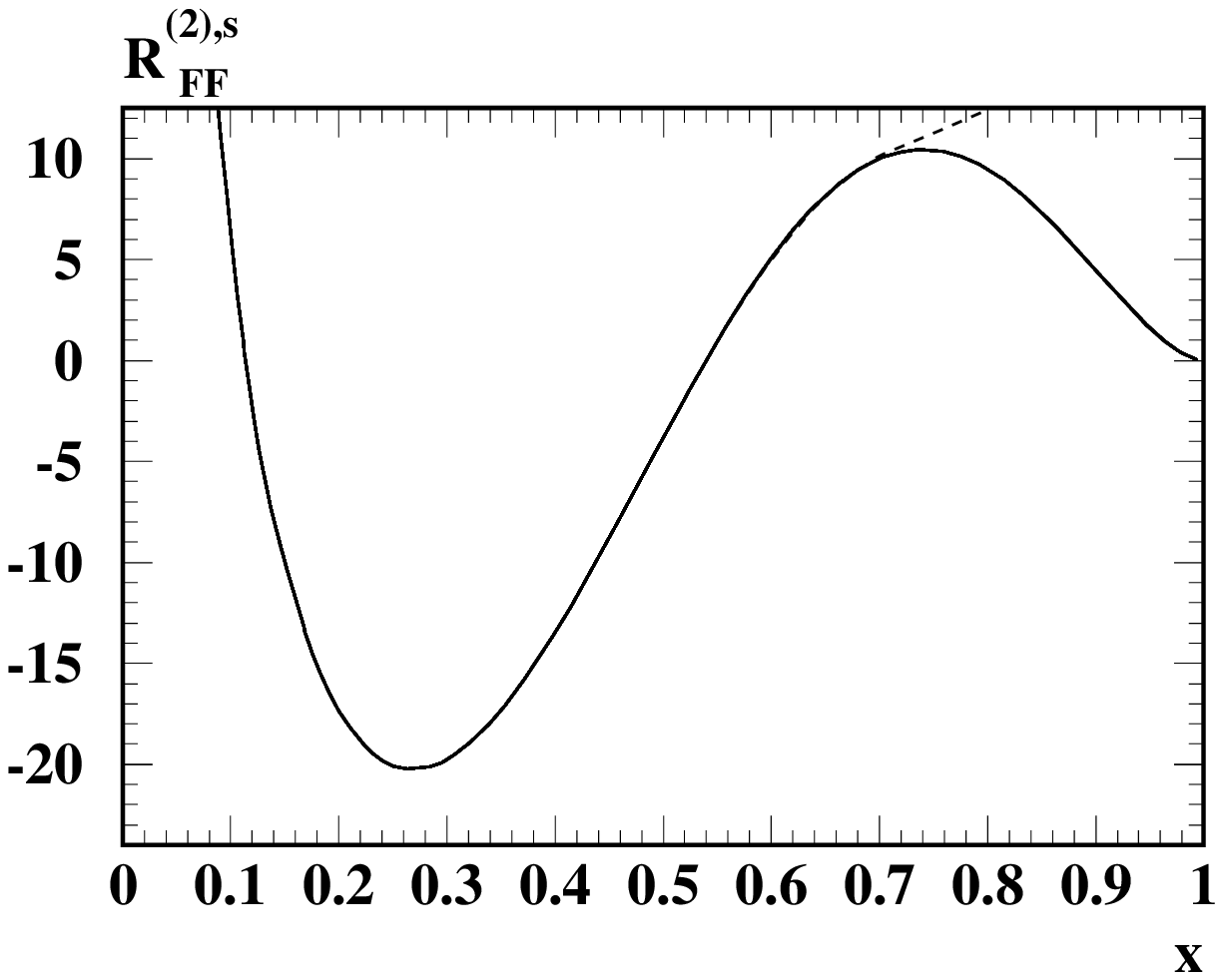}
    \end{tabular}
  \end{center}
  \caption{\label{fig:Rvx_cf2}$R^{(2),v}_{FF}(s)$ and $R^{(2),s}_{FF}(s)$
    as a function of $x$ (full line). The dashed curves correspond to
    the high-energy results obtained via asymptotic expansion 
    including terms up to order $1/z^7$.  
          }
\end{figure}

In order to get even more confidence in the results
we plot in Fig.~\ref{fig:Rvx_cf2} 
$R^{(2),v}_{FF}(s)$ and $R^{(2),s}_{FF}(s)$ as a function of $x$.
Expressed in this variable the high energy region gets more spread. Thus 
a sensible comparison with the expansion terms can be performed.
The dashed curves correspond to the analytical 
expressions obtained via asymptotic expansion containing
the terms up to order $1/z^7$. Excellent agreement with the
semi-numerical results is observed up to 
$x\approx 0.5$ for the vector case and $x\approx 0.7$ for the scalar
case which corresponds to $v\approx 0.60$ and $v\approx 0.34$, respectively.
This is a strong consistency check for the Pad\'e method.

\subsection*{Single-top-quark production}

\begin{figure}[t]
  \begin{center}
    \begin{tabular}{c}
      \leavevmode
      \epsfxsize=14.cm
      \epsffile[70 580 516 720]{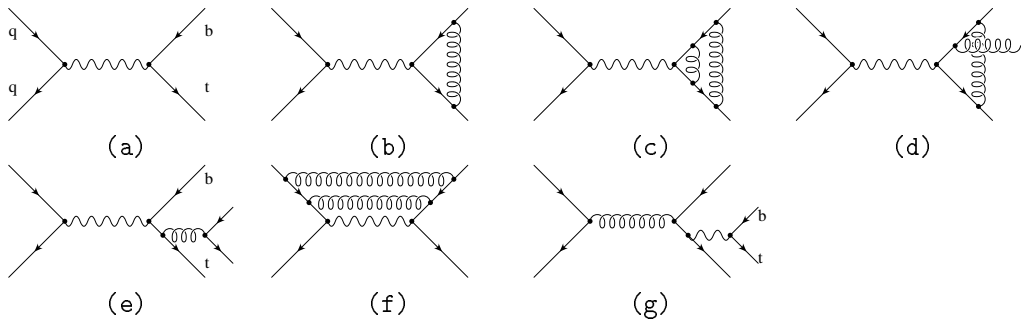}
    \end{tabular}
  \end{center}
  \caption{\label{fig:diags_qqtb}Sample diagrams contributing to the
    process $q\bar{q}\to t\bar{b}$. The wavy and loopy lines 
    represent $W$ bosons and gluons, respectively.
          }
\end{figure}

As a first application we want to discuss the single-top-quark production
via the process $q\bar{q}\to t\bar{b}$. Some sample diagrams are
plotted in Fig.~\ref{fig:diags_qqtb}.

The corrections of order $\alpha_s$  to the (total) single-top-quark 
production rate are quite large. They amount to
about  54\% and 50\% for Tevatron and LHC energies,
respectively~\cite{SmithWillen96}, where 18\%, respectively, 17\% arise
from the final state corrections.
This calls for a complete   ${\cal O}(\alpha_s^2)$ calculation.
In this letter we want to do a first step and consider the leading
term in the large-$N_c$ expansion.

At ${\cal O}(\alpha_s)$ there is no interference between the initial
and final state radiation. Thus it is possible to write 
the differential cross section is factorized form
\begin{eqnarray}
  \frac{{\rm d}\sigma}{{\rm d}q^2} (p\bar{p} \to t\bar{b}+X)
  &=& \sigma(p \bar{p} \to W^* +X) 
  \frac{\mathrm{Im}\, \Pi_W(q^2,M_t^2,M_b^2)}{\pi (q^2-M_W^2)^2}
  \label{eq:diff_cross}
  \,,
\end{eqnarray}
where $\Pi_W$ corresponds to the transversal part of the $W$ boson
correlator and is connected to the vector correlator of
Eq.~(\ref{eq:pivadef}) through
\begin{eqnarray}
  \Pi_W(q^2) &=& \sqrt{2} G_F M_W^2 |V_{tb}|^2 q^2 \Pi^v(q^2)
  \,.
  \label{eq:piw}
\end{eqnarray}
Due to interference diagrams between the initial and final state 
(c.f. Fig.~\ref{fig:diags_qqtb}(f))
Eq.~(\ref{eq:diff_cross}) does not hold at $\alpha_s^2$. 
However, those contributions are
suppressed by at least a factor $1/N_c^2$ in large $N_c$ limit
as compared to the diagrams in Fig.~\ref{fig:diags_qqtb}(c)--(e).
The latter, together with the contributions in 
Fig.~\ref{fig:diags_qqtb}(a) and~(b),
are covered by Eqs.~(\ref{eq:diff_cross}) and~(\ref{eq:piw}).

At order $\alpha_s^2$ there are also diagrams like the one in
Fig.~\ref{fig:diags_qqtb}(g) which appear for the first time.
In principle they lead to the same final state as 
the diagram in Fig.~\ref{fig:diags_qqtb}(e). However, one 
has to note that the $W$ boson generating the top and bottom quark
is radiated from a light quark flavour.
Furthermore, compared to the diagram in Fig.~\ref{fig:diags_qqtb}(c)
these contributions are only suppressed by a factor $1/N_c$
and not by $1/N_c^2$ like the diagram in Fig.~\ref{fig:diags_qqtb}(f).

Thus, if we restrict ourselves to the leading term in $1/N_c$
it is possible to use the results for $R^v$ 
obtained above in combination
with Eq.~(\ref{eq:diff_cross}) to perform a theoretical analysis of
the order $\alpha_s^2$ to the single-top-quark production in the
large-$N_c$ limit. In order to obtain the total cross section the 
corresponding parton distribution functions would be needed to the
same order.

\begin{figure}[t]
  \begin{center}
    \begin{tabular}{c}
      \leavevmode
      \epsfxsize=14.cm
      \epsffile[110 280 470 560]{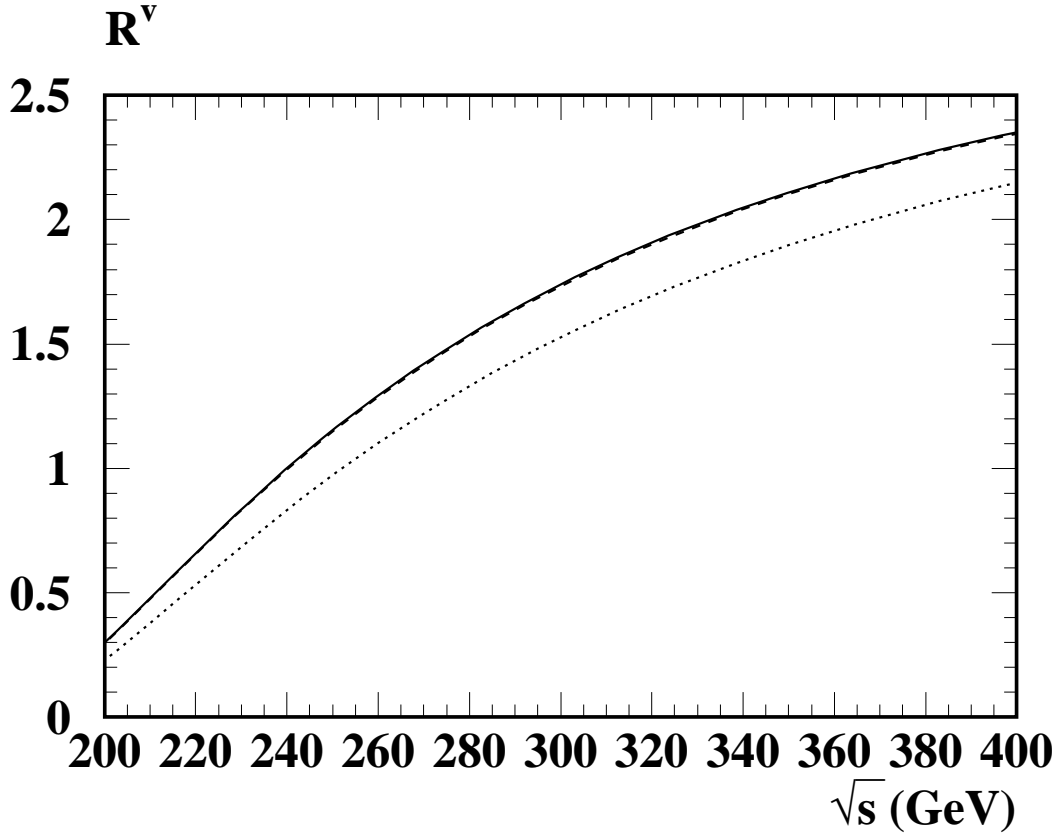}
    \end{tabular}
  \end{center}
  \caption{\label{fig:dsigma_qqtb}LO (dotted), 
    NLO (dashed) and NNLO (solid) results of $R^v(s)$.
          }
\end{figure}

The production cross-section of the virtual $W^*$ boson is identical to
that of the Drell-Yan process $q\bar{q}\to e\bar{\nu}_e$.
The latter is known to
${\cal{O}}(\alpha^2_s)$ from Ref.~\cite{Drell_Yan}.
Thus we can take the proper ratios to make predictions 
in the large-$N_c$ limit 
at NNLO free from any  dependence on parton distribution functions.  
As an example, we consider
\begin{eqnarray}
  \frac{\frac{{\rm d}\sigma}{{\rm d}q^2}\left(pp\to W^*\to tb\right)}
       {\frac{{\rm d}\sigma}{{\rm d}q^2}\left(pp\to W^*\to
       e\nu_e\right)}
  &=& \frac{\mbox{Im}\left[\Pi_{tb}(q^2)\right]}
           {\mbox{Im}\left[\Pi_{e\nu}(q^2)\right]} 
  \nonumber\\
  &=& N_c |V_{tb}|^2 R^v(s)
  \,.
  \label{eq:ratio}
\end{eqnarray}
To get an impression of the numerical significance we plot
in Fig.~\ref{fig:dsigma_qqtb} the LO, NLO and NNLO result
of $R^v(s)$ in the range $\sqrt{s} = 200 \ldots 400$ GeV. 
For the numerical values we choose $M_t=175$~GeV and 
$\alpha_s(M_Z)=0.118$.
Whereas the ${\cal O}(\alpha_s)$ corrections are significant
there is only a moderate contribution from the order $\alpha_s^2$
terms. In the range in $q^2$ shown in Fig.~\ref{fig:dsigma_qqtb} they are
below 1\% of the Born result.
Note that the NNLO correction to the Drell-Yan process are also 
small and amount to at most a few percent (see e.g.~\cite{Mar00}).
Thus, in case there is no kinematical magnification for the 
diagrams in Fig.~\ref{fig:diags_qqtb}(f) and~(g) we can conclude
that the radiative corrections to the single-top-quark production via the 
process $q\bar{q}\to t\bar{b}$ are well under control.

\subsection*{Decay rate of a charged Higgs}

Theories beyond the SM are usually characterized by an enlarged Higgs
sector and may allow for different quantum numbers of the Higgs
bosons.  For example, one of the most appealing extensions of the SM,
the Minimal Supersymmetric Standard Model (MSSM), contains two complex
iso-doublets with opposite hyper-charge (see
e.g.~\cite{GunHabKanDaw90}), resulting in five mass eigenstates of
(pseudo-)scalar physical Higgs fields: 
two neutral CP-even ($H^0$ and $h^0$),
one neutral CP-odd ($A$)
and two charged ($H^{\pm}$) Higgs bosons.

Let us consider a generic charged Higgs boson coupled to 
fermions through
\begin{eqnarray}
  {\cal L}_{H^+D\bar{U}} &=&
  \left(\sqrt{2} G_F\right)^{1/2} \, H^+\, J_{H^+}
  \,,
\end{eqnarray}
where the corresponding quark current is given by
\begin{eqnarray}
  J_{H^+} &=& 
  \frac{m_U}{\sqrt{2}} \,  
  \bar{U} \left[ a \, (1-\gamma_5) + b \,  (1+\gamma_5) \right] D
  \label{eq:Higgs_current} 
  \,.
\end{eqnarray}
Here $U$ and $D$ represent generic up- and down-type quarks,
respectively, with $\overline{\rm MS}$ masses $m_U$ and $m_D=0$.
We only consider $H^+$ as for $H^-$ the formulae are analogues.
The parameters $a$ and $b$ are model dependent and are left
unspecified.

\begin{figure}[t]
  \begin{center}
    \begin{tabular}{c}
      \leavevmode
      \epsfxsize=14.cm
      \epsffile[110 280 470 560]{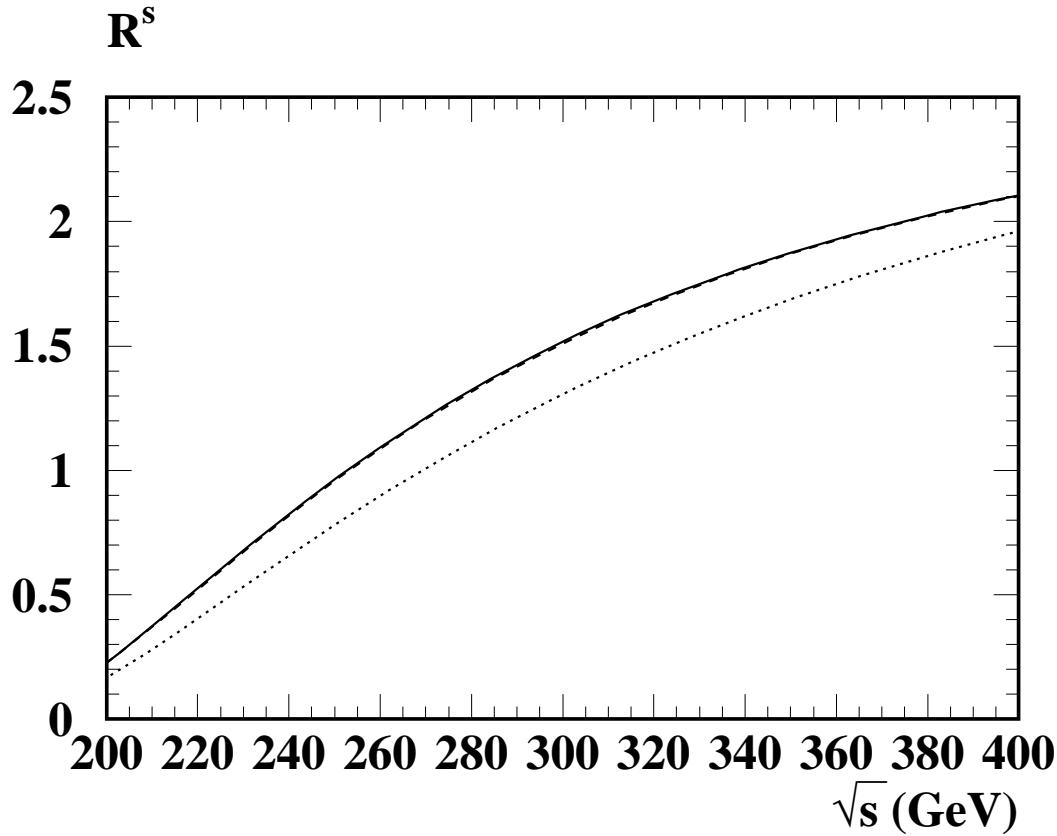}
    \end{tabular}
  \end{center}
  \caption{\label{fig:higgs}LO (dotted), 
    NLO (dashed) and NNLO (solid) results of $R^s(s)$, $M=M_t=175$~GeV.
          }
\end{figure}

The  decay rate of the Higgs boson $H^+$ into quarks and gluons
can  be written in the form 
\begin{eqnarray}
  \Gamma(H^+ \to U \bar{D})
  &=& 
  \sqrt{2} G_F M_{H^+} \, \mathrm{Im}\, \left[ \Pi_H(M_{H^+}^2) \right]
  \,,
\end{eqnarray}
where $M_U$ is the pole quark mass and $\Pi_H(q^2)$ is given by
\begin{eqnarray}
  q^2 \Pi_H(q^2) &=& \int {\rm d}x \, e^{iqx} 
  \langle     
  T J_H{^+}(x) J_{H^-}(0)
  \rangle
  \,\,=\,\,(a^2+b^2) q^2 \Pi^s(q^2) 
  \label{eq:Higgs_corr}
  \,,
\end{eqnarray}
Thus, we arrive at the following expression for
the hadronic decay rate of the charged Higgs boson
\begin{eqnarray}
  \Gamma(H^+ \to U \bar{D})
  & = &\frac{\sqrt{2}G_F}{8\pi}M_{H^+} (a^2+b^2) R^s(M_{H^+}^2)
  \,.
\end{eqnarray}
In Fig.~\ref{fig:higgs} $R^s(M_{H^+}^2)$ is plotted at LO, NLO and NNLO.
Again it turns out that the radiative corrections are well under
control as order $\alpha_s^2$ terms contribute at most of the order of 1\%.

\section{ \label{sec:con} Conclusions }

We have computed the non-diagonal correlators of vector and scalar
currents at order $\alpha_s^2$ in the case of the currents composed
from a massless and a massive quark field. Our main theoretical tools
have been the asymptotic expansions in the large and small momentum
region as well
as some information about the threshold behaviour provided by HQET.

The obtained results for the case of the (axial-)vector correlator
constitute an important ingredient of the full calculation of
next-to-next-to-leading ${\cal O}(\alpha_s^2)$ correction to the
single-top-quark production via the process $q\bar{q}\to t\bar{b}$.
Such a calculation is required for an accurate extraction of
the matrix element $|V_{tb}|$ from experiment.

We have used the results for the (pseudo-)scalar correlator in order
to obtain the ${\cal O}(\alpha_s^2)$ decay rate of a charged (pseudo-)scalar
Higgs boson to hadrons.


\section*{Acknowledgments}

The authors are grateful to A. Grozin, J.H. K\"uhn and V.A. Smirnov 
for useful discussions and advice.  

This work was supported in part by the {\it DFG \ DFG-Forschergruppe
``Quantenfeldtheorie, Computeralgebra und Monte-Carlo-Simulation''} 
(contract FOR 264/2-1) and by SUN Microsystems through Academic
Equipment Grant No.~14WU0148.


\end{document}